Title: Interoception Underlies The Therapeutic Effects of Mindfulness Meditation for Post-Traumatic Stress Disorder: A Randomized Clinical Trial

Seung Suk Kang, Ph.D.[1*], Scott R. Sponheim, Ph.D.[3,2], Kelvin O. Lim, M.D.[2,3]

[1]Department of Biomedical Sciences, University of Missouri-Kansas City, Kansas City, MO, 64108, USA
[2]Department of Psychiatry, University of Minnesota, Minneapolis, MN, 55455, USA
[3]Veterans Affairs Health Care System, Minneapolis, MN, 55417, USA

Corresponding Author: Seung Suk Kang, Ph.D. kangseung@umkc.edu

## ABSTRACT

**IMPORTANCE**: Mindfulness-based interventions have proven its efficacy in treating post-traumatic stress disorder (PTSD), but the underlying neurobiological mechanism is unknown.

**OBJECTIVE**: To determine the neurobiological mechanism of action of mindfulness-based stress reduction (MBSR) treating PTSD.

**DESIGN, SETTING, AND PARTICIPANTS**: Randomized clinical trial of 98 veterans with PTSD recruited at the Minneapolis Veterans Affairs Medical Center (MVAHS) from March 2012 to December 2013. Outcomes were assessed before and after treatment. Data collection was completed on April 22, 2014.

**INTERVENTION**: Participants were randomly assigned to receive MBSR therapy (n = 47), consisting of 9 sessions (8 weekly 2.5-hour group sessions and a daylong retreat) focused on teaching patients to attend to the present moment in a nonjudgmental, accepting manner; or present-centered group therapy (PCGT; n = 51), an active-control condition consisting of 9 weekly 1.5-hour group sessions focused on current life problems.

**MAIN OUTCOMES and MEASURES**: PTSD Checklist was used to assess the primary outcome, change in PTSD symptom severity at baseline and post-intervention. Secondary outcomes were assessed using brain activity measures of electroencephalography (EEG), including spectral power of spontaneous neural oscillatory activities during resting and meditation periods, time-frequency power of cognitive task-related brain responses, and time-frequency power of heartbeat-evoked brain responses (HEBR) that reflect cardiac interoceptive brain responses during resting and meditation.

**RESULTS**: Compared to controls, the MBSR group had greater improvements in PTSD symptoms, spontaneous EEG alpha (8-13 Hz) power in posterior sites, task-related frontal theta power (4-7 Hz in 140-220 ms post-stimulus), and frontal theta HEBR (3-5 Hz and 265-328 ms post-R-peak). Latent difference score modeling found that only the changes in the frontal theta HEBR mediated the MBSR treatment effect. Brain source-level analysis found that the theta HEBR changes in the anterior cingulate cortex, anterior insular cortex, and the lateral prefrontal cortex predicted PTSD symptom improvements.

**CONCLUSIONS AND RELEVANCE**: These results indicated that mindfulness meditation improves spontaneous brain activities reflecting internally oriented relaxation and brain functions of attentional control. However, interoceptive brain capacity enhanced by MBSR appears to be the primary cerebral mechanism that regulates emotional disturbances and improves anxiety symptoms of PTSD.

**TRIAL REGISTRATION**: clinicaltrials.gov Identifier: NCT01548742

## INTRODUCTION

Mindfulness meditation is a mind-body practice that cultivates nonjudgemental attention to present-moment experiences, enhances various psychological functions, and reduces stress[1–3]. Various mindfulness-based approaches have been developed for treating chronic pain, depression, and anxiety spectrum disorders[4,5]. In particular, mindfulness-based stress reduction (MBSR) has been recognized as an alternative intervention for severe psychopathology that conventional interventions have shown only limited success, such as post-traumatic stress disorders (PTSD)[6–8]. PTSD is a potentially debilitating mental disorder characterized by re-experience of traumatic events, hyper-arousal, and avoidance symptoms. Recently we demonstrated that MBSR had superior efficacy treating PTSD symptoms compared to an active control intervention[7]. However, the precise underlying neurobiological mechanism of action is unknown.

Mindfulness meditation enhances self-regulation by improving attention-control, emotional regulation, and altered self-awareness that is characterized by diminished self-referential thinking and enhanced body awareness[1]. An activation likelihood estimation meta-analysis identified multiple brain regions whose structures were consistently altered in meditators across studies[9], including those related to attention-control (the rostral, lateral, and inferior prefrontal cortex [PFC]), self- and emotion-regulation (anterior cingulate cortex [ACC]), and body awareness (the somatosensory cortex and insula). Numerous neurophysiological studies also have found various state and trait changes in the brain activities related to meditation[10]. A systematic review of electroencephalography (EEG) studies of mindfulness meditation identified the most consistent findings including 1) increases in the theta (3-7 Hz) and alpha (8-13 Hz) frequency spontaneous oscillations during resting and meditation that might reflect a mental state of relaxed alertness and 2) attentional cognitive task-related brain activity improvements[11].

With heightened recognition of the integral role of interoception in emotional experience, self-regulation, decision making, and consciousness[12,13], recent studies theorized that beneficial effects of mindfulness meditation arise from cultivating interoceptive abilities[14–16]. Interoception refers to processes by which the body senses, interprets, integrates, and regulates signals from within itself, which represents the core component of homeostatic processes[17,18]. Behavioral and neuroimaging studies found that mindfulness meditation practitioners had high performance accuracy[19] and increased brain responses, especially in insula[14] for respiratory interoceptive sensation tasks. However, it is unknown if the functional changes in the spontaneous, cognitive, and interoceptive brain activities mediate the therapeutic effects of mindfulness meditation.

To determine the neural mechanism of action of MBSR, we conducted a parallel design mechanistic clinical trial in which veterans with PTSD were randomly assigned to 8-week programs of MBSR or an active control intervention and assessed for clinical and neurophysiological outcomes. Spontaneous and task EEG recordings were collected with a resting-meditation-resting procedure and an attentional cognitive task. We assessed spectral power of the theta and alpha frequency oscillations of the spontaneous EEG and time-frequency (TF) power of the task-related oscillatory brain responses. Also, we assessed TF-power measures of heartbeat-evoked brain responses (HEBR), the most widely used interoceptive brain response measurements that are known to reflect interoceptive attention[20–22], emotion[23–25] and self-related neural processes[26,27]. With EEG measures showing significant MBSR-related changes, we conducted causal mediation analyses to test if they mediated the therapeutic effects of MBSR. We hypothesized that MBSR would increase HEBR, which would have a significant mediation effect.

## METHODS

### Participants, Treatment Conditions, and Procedures

The institutional review board of MVAHS approved this study. After providing informed consent, participants completed an eligibility assessment using a structured clinical interview and self-report measures, including the PTSD Checklist[28] (PCL) that was the primary clinical outcome. Demographics and the baseline clinical characteristics of the participants are summarized in Table 1. MBSR involving body scanning, sitting meditation, and mindful yoga practices and present-centered group therapy (PCGT)[29,30] were used to control for nonspecific therapeutic factors[31]. Eight-week treatment was delivered in a group format according to manualized protocols by 2 instructors/clinicians. We randomized participants approximately every 2 months over a 19-month period, for a total of 9 cohorts composed of 1 group each of the 2 conditions. Participants completed one-hour EEG recording sessions at baseline and post-intervention. Only cohort 1-8 had EEG recordings. The details of the participant recruitment, inclusion/exclusion criteria, treatment conditions, procedures of clinical outcome evaluation, randomization, and blinding are described in Polusny et al.[7]

**Table 1. Demographic Characteristics and the Baseline Clinical Assessment Scores of the Participants**

| Demographic Characteristics | Total | MBSR (n=47) | PCGT (n=51) |
|---|---|---|---|
| Male | 84 | 38 | 46 |
| Female | 14 | 9 | 5 |
| Age | 59.0 (9.8) | 58.6 (10.4) | 59.4 (9.3) |
| Race | | | |
|   White | 81 | 37 | 44 |
|   Black | 8 | 3 | 5 |
|   Other | 4 | 2 | 2 |
|   Mixed | 5 | 5 | 0 |
| Baseline clinical assessment scores, mean (SD)c | | | |
| Self-reported PTSD symptom severity on the PCL | 60.8 (12.0) | 62.2 (10.42) | 59.6 (13.3) |
| Interview-rated PTSD severity on the CAPS | 65.5 (16.3) | 68.1 (15.0) | 63.2 (17.2) |
| Self-reported depression symptom severity on the PHQ-9 | 15.2 (5.2) | 15.3 (4.9) | 15.2 (5.4) |
| Self-reported somatic symptom severity on the PHQ-15 | 13.5 (4.8) | 13.8 (4.3) | 13.1 (5.2) |

*Note*: No significant group difference was observed in any demographic variables and the baseline clinical assessment scores. *Abbreviations*: CAPS, Clinician-Administered PTSD Scale; MBSR, Mindfulness-Based Stress Reduction; PCGT, Present-Centered Group Therapy; PCL, PTSD Checklist; PHQ-9, Patient Health Questionnaire 9; PTSD, posttraumatic stress disorder;

**Electroencephalography (EEG) and Electrocardiogram (ECG) Recording Procedure**

EEG was recorded using a Biosemi Active Two EEG system (http://www.biosemi.com) with 64 Ag/AGcl electrodes aligned with the 10-10 international system. ECG was acquired using two Ag/Agcl electrodes placed on the left and right clavicles. EEG and ECG signals were digitized at a 1024 Hz sampling rate and online referenced to the BioSemi CMS-DRL (common mode sense-driven right leg) reference[32]. Each EEG recording session began with 4 minutes resting with two sets of alternating one-minute eyes-closed (EC) and one-minute eyes-open (EO) resting-state recordings (EC-EO-EC-EO). Then, EEG data was recorded for 10 minutes during eyes-closed meditation. To test if resting-state EEG activities were changed after meditation, we also recorded another 4 minutes resting (EO-EC-EO-EC) EEG session. PCGT participants were asked to rest with their eyes closed (REC) during the meditation period. In baseline EEG sessions prior to MBSR instructions, MBSR participants were also asked to rest with their eyes closed during the meditation period. Fatigue and sleepiness were checked at each EEG session. Lastly, participants performed a flanker cognitive task[33], in which they watched a computer screen displaying visual stimuli and were instructed to press a right or left button according to the direction of the central arrow regardless of the directions of the flanker arrows.

**EEG and ECG Data Processing**

EEG and ECG data were preprocessed using custom Matlab-based programs, which performed digital filtering (pass-band: .5-128 Hz) and downsampling (256 Hz). Resting and meditation EEG recordings were segmented into 2-second epochs while flanker task EEG was segmented for 500 ms pre-stimulus and 1500 ms post-stimulus time-windows. Independent component analysis (ICA) that employed the Fast ICA algorithm[34,35] was used to remove non-brain signal artifacts from the EEG. Epochs with atypical artifacts from body or electrode movements and muscle contractions were rejected before ICA decomposition of the EEG signals and identified through the presence of prominent low (<8 Hz) or high (>25 Hz) frequency artifacts observed on visual inspection. Unlike stereotyped artifacts such as eye-blinks or electrical line noises, these noisy time-segments indicate non-stereotyped artifacts compromising ICA decomposition[36]. To obtain an optimal ICA decomposition, preventing under- or over-fitting of the EEG data that results in ICs with mixed brain and non-brain signals, we carried out principal component analysis (PCA)[37]. PCA conducted prior to ICA reduced the original data dimensionality to its intrinsic dimensions. The intrinsic dimensionality was estimated by a Bayesian model order selection method based on maximum likelihood on the eigenvalues of EEG data[38]. ICs identified as capturing signal artifacts were removed from the EEG data. ICA-denoised EEG was re-referenced to the average head.

As Park and Blanke[39] discussed, there are distinct methodological issues in analyzing HEBR of EEG signals time-locked to ECG R-peaks that are largely affected by physiological noise, including cardiac field artifact (CFA)[40] and pulse artifact[41]. It is extremely hard to completely remove these cardiac artifacts using any denoising technique, including ICA approach[22]. To prevent any individual differences in HEBR measures caused by varying degrees of remaining ECG artifacts due to incomplete decompositions of ECG artifact ICs, we did not remove any ICs with ECG artifacts during EEG preprocessing. Instead, we contrasted HEBR between the pre- and post-intervention EEG (W9–W0) to quantify the intervention-related HEBR changes that were minimally affected by the cardiac artifacts. ECG data was filtered further (low-pass filtering with 45 Hz cut-off frequency) to reduce high-frequency noise that could lead to inaccurate detection of ECG R-peaks. R-peak time-points of ECG data were identified using a custom Matlab-based software that detects ECG R-peaks based on an amplitude threshold and the velocity of ECG peaks. Every identified R-peak was visually inspected, and any erroneous peaks were manually corrected.

**Spontaneous EEG Oscillatory Activities**
Spontaneous oscillatory brain activity during resting and meditation states were quantified using a spectral power analysis by applying Fast-Fourier Transforms (FFTs) to two-second epochs tapered with Hamming window to resolve frequency spectra for 64 electrodes for each subject. The frequency power ($\mu V^2$) was log-transformed to yield rectified distributions close to normal. The trial-level EEG spectra were averaged for the first EC resting (Rest-1), meditation (or 10-min REC), the second EC resting (Rest-2), and the entire EC EEG recordings to test state-related EEG spectra differences. Given the obviously different EEG spectra between EO and EC states[42], EO Resting EEG spectra were not analyzed in the current study.

**Flanker Cognitive Task-Related Brain Responses**
EEG time-locked to the visual stimuli onsets were time-frequency (TF) transformed using Cohen's class TF-transformation algorithm[43] with reduced interference distribution (RID) method that provides uniform time-frequency resolution[44] using a binomial time-frequency distribution (TFD) kernel[45]. The cognitive task-related EEG oscillatory power TFDs were averaged for the congruent and incongruent directional flanker conditions and normalized as percent signal changes using the mean TF-power for each frequency in a pre-stimulus baseline time-window (-300 to 0 ms).

**Heartbeat-Evoked Brain Responses (HEBR)**
To measure HEBR, resting and meditation EEG data were segmented from -500 ms to 850 ms relative to R-peaks. We processed and analyze HEBR to minimize the influences of cardiac filed artifacts and pulse artifacts in HEBR measures as Park and Blanke suggested[39], focusing on the difference between the pre- and post-intervention EEG (W8–W0) to quantify the intervention-related HEBR changes. To delineate the time and frequency properties of the intervention-related subtle changes in oscillatory HEBR, we computed the TF-power of EEG time-locked to R-peaks using the methods described above. The heartbeat-related EEG oscillatory power TFDs were averaged for resting-1, meditation (or REC), resting-2, and the entire EC recordings. To quantify transient oscillatory HEBR changes compared to the pre-R-peak time-window, we computed normalized HEBR TFDs as percent signal changes compared to the mean TF-power for each frequency within a pre-R-peak baseline time-window (-300 to -100 ms)[46]. The baseline time-window was chosen to avoid 1) possible influence of cardiac artifacts around the ECG Q-R-S peak complex[47] whose amplitudes and TF-power were somewhat different between the pre- and post-intervention ECG (see Figure 4A), and 2) possible overlap between the pre-R-peak baseline and the post-R-peak time-windows of interest.

**Statistical Analyses**
For statistical analyses of the spontaneous EEG oscillations, we quantified the mean log power of theta (3-7 Hz) and alpha (8-13 Hz) frequency of 64-channel EEG in the 3 states (Rest-1, meditation [or REC], and Rest-2) that have been consistently reported to have meditation-related effects[11]. To prevent circular inference[48], we identified the TF-windows showing intervention-related changes in the HEBR and the task-related power TFDs in the W8-W0 difference grand average TFDs collapsed across the two groups and the three states or the two task conditions. The mean powers of the identified TF-windows in 64 electrodes were computed for the 3 states or the 2 task conditions for statistical analyses. Using SPSS software version 24.0 (IBM Corp), we conducted repeated-measure ANOVA (RM-ANOVA) for each EEG measures with a between-group factor of the intervention type (MBSR and PCGT) and 3 within-group factors including time (W0 and W8), states (for the spontaneous and HEBR EEG measures: Rest-1, meditation [or REC], and Rest-2) or task conditions (for the task EEG measure: congruent and incongruent), and electrodes (64-channels). Hyunh-Feldt correction was used when a sphericity assumption was violated[49]. Follow-up paired, or independent *t*-tests were used to examine the interaction effects found in the RM-ANOVA.

To determine the scalp electrodes having significant intervention-related changes, we conducted electrode-cluster based permutation tests[50] using the following steps. First, we identified electrode-clusters as groups of 3 or more adjacent electrodes with significant time (W8-W0) effects in paired *t*-tests in each group or those with significant group effects in independent *t*-tests of the W0, W8, or W8-W0 difference EEG measures. Second, we permuted time or group 10,000 times and computed the sum of the *t*-values of the electrode-clusters in the corresponding *t*-tests with each permutation dataset to generate null-hypothesis distributions of the differences in the electrode-cluster measures. Third, the statistical significance of the intervention-related EEG measures of the electrode-clusters was tested based on the permutation *p*-values (i.e., proportions of the permutation *t*-tests whose absolute *t*-values exceed the absolute *t*-values of the original *t*-tests).

**Mediation Analyses**

To test if changes in the EEG measures mediated the therapeutic effect of MBSR, we conducted mediation analyses using latent difference score (LDS) modeling[51,52], a specific subtype of longitudinal structural equation modeling (SEM). Addressing the auto-regressiveness of the measurements, LDS modeling can estimate latent difference scores (i.e., changes in the latent variable) from longitudinal measurements, even with measures from only two time points[53]. We included three sets of variables in our LDS models, including the predictor of the intervention type (X), and the clinical outcome variable of PTSD symptom severity (the sum of PCL score; Y), and a mediator of one of the EEG measures (M) showing significant group-by-time interaction effects. The mediator EEG measures were quantified as the mean of the EEG measures in the electrode clusters showing the group-by-time interaction effects. Using Mplus[54] version 8, we estimated the LDS of the outcome and the mediator ($\Delta_M$ and $\Delta_Y$) and tested the direct effect of X and the indirect (mediation) effect of X through $\Delta_M$ on Y (see Figure 5A). The statistical significance of the effects was determined based on the 95% confidence intervals of the path coefficients. The Mplus syntax used in the present mediation analysis is available below.

**Brain Source Imaging of the EEG Measure**

As the frontal theta HEBR changes mediated the therapeutic effect of MBSR, we localized the brain sources of the theta HEBR whose changes predicted the PTSD symptom reductions. We used a beam-forming based spatial filter technique called dynamic imaging of coherent sources (DICS)[55] and Fieldtrip Matlab toolbox[56] (see eFigure 4 for the procedure). For the forward modeling of the brain source localization, we used the standard MNI/ICBM152 brain[57] to create finite element method (FEM) volume conduction model of the head, which models the structure and conductivity of five compartments of the brain (skin, skull, CSF, gray matter, and white matter; conductivity: 0.43, 0.0024, 1.79, 0.14, and 0.33, respectively). The 10-10 international system standard montage of the 64-channel electrodes were fitted to the head model. For the inverse modeling of the 64-channel TF-power measures, we calculated the cross-spectral density matrix of the 3-5 Hz oscillatory activities in 64-channel electrodes. We computed neural activity indices (NAI)[58] of the baseline (-300 to -100 ms pre-R-peak) and the HEBR time-windows (265-328 ms post-R-peak) by dividing the power estimates by the noise estimates. Lastly, the brain maps of the baseline-normalized NAI (%Change) were calculated for statistical analyses.

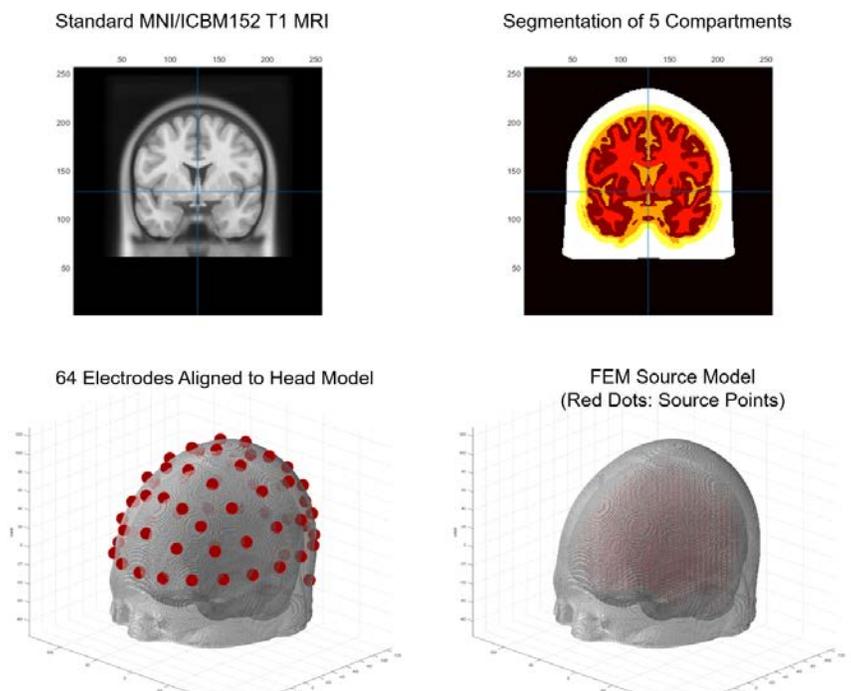

**Figure 1. Finite element method (FEM) Head Model and 64-Channel EEG Registered to the Head Model.**

# RESULTS

**Clinical Outcome**

Figure 2 depicts participant flow through the study. All participants had PTSD symptom improvements after 8-week interventions, but a significant group-by-time interaction was observed ($F[1,96]=9.52$, $p=.003$, $\eta^2=.090$), which indicated that there was a greater improvement in PTSD symptoms in the MBSR than the PCGT group (mean PCL score improvement: MBSR from 62.2 to 54.4 vs. PCGT from 59.6 to 57.9; mean difference in improvement, 5.07; 95% CI: 2.15-9.0; $t[87]=2.57$; $p=.012$; $d=.624$).

**Spontaneous EEG Power**

The theta (3-7 Hz) power of spontaneous EEG recordings had a trend level increase after intervention (time effect: $F[1,87]=3.25$, $p=.075$, $\eta^2=.036$), but there was no group difference in the increase (group-by-time effect: $F[1,87]=1.46$, $p=.230$). Analysis of the alpha (8-13 Hz) power across all scalp recording sites found a trend level group-by-time interaction effect ($F[1,87]=3.78$, $p=.055$, $\eta^2=.042$) and significant state ($F[1,174]=7.75$, $p=.002$, $\eta^2=.072$) and electrode ($F[63,5481]=150.93$, $p<10^{-100}$, $\eta^2=.634$) effects. Electrode cluster-based permutation tests to investigate effects for select recording sites revealed that only MBSR group had significant alpha power increase in the posterior electrode cluster ($p=.026$, $d=.33$). The group difference in the intervention-related alpha power change was significant in the right posterior electrode cluster ($p=.005$, $d=.55$). Follow-up analyses of the mean alpha power of the posterior electrodes showing the group-by-time interaction effects found that each group had similar intervention-related changes across the 3 states (Figure 3C).

**Cognitive Task-Related Brain Responses**

Flanker task performances were improved in both groups after intervention, showing time effects in %Correct ($F[1,86]=12.99$, $p=.001$, $\eta^2=.131$) and the mean reaction time (MRT; $F[1,86]=4.29$, $p=.041$, $\eta^2=.048$). However, group-by-time interaction effects were not significant in both measures (%Correct: $F[1,86]=1.64$, $p=.204$; MRT; $F[1,86]=.72$, $p=.397$). The pre- vs. post-intervention difference TFDs showed substantial increases in the early theta power (4-7 Hz in 140-220 ms after stimulus onsets) in the combined group. RM-ANOVA of the theta component found a significant group-by-time interaction effect ($F[1,86]=6.60$, $p=.012$, $\eta^2=.071$). Only MBSR group had a significant increase in the theta component in the frontal electrode cluster ($p<.001$, $d=.39$; Figure 5). The group difference in the intervention-related theta component change was significant in the similar frontal electrode cluster ($p<.001$, $d=.54$). Follow-up analyses of the mean theta power of the frontal electrodes showing the group-by-time interaction effects found that each group had similar intervention-related changes between the 2 conditions (Figure 4E).

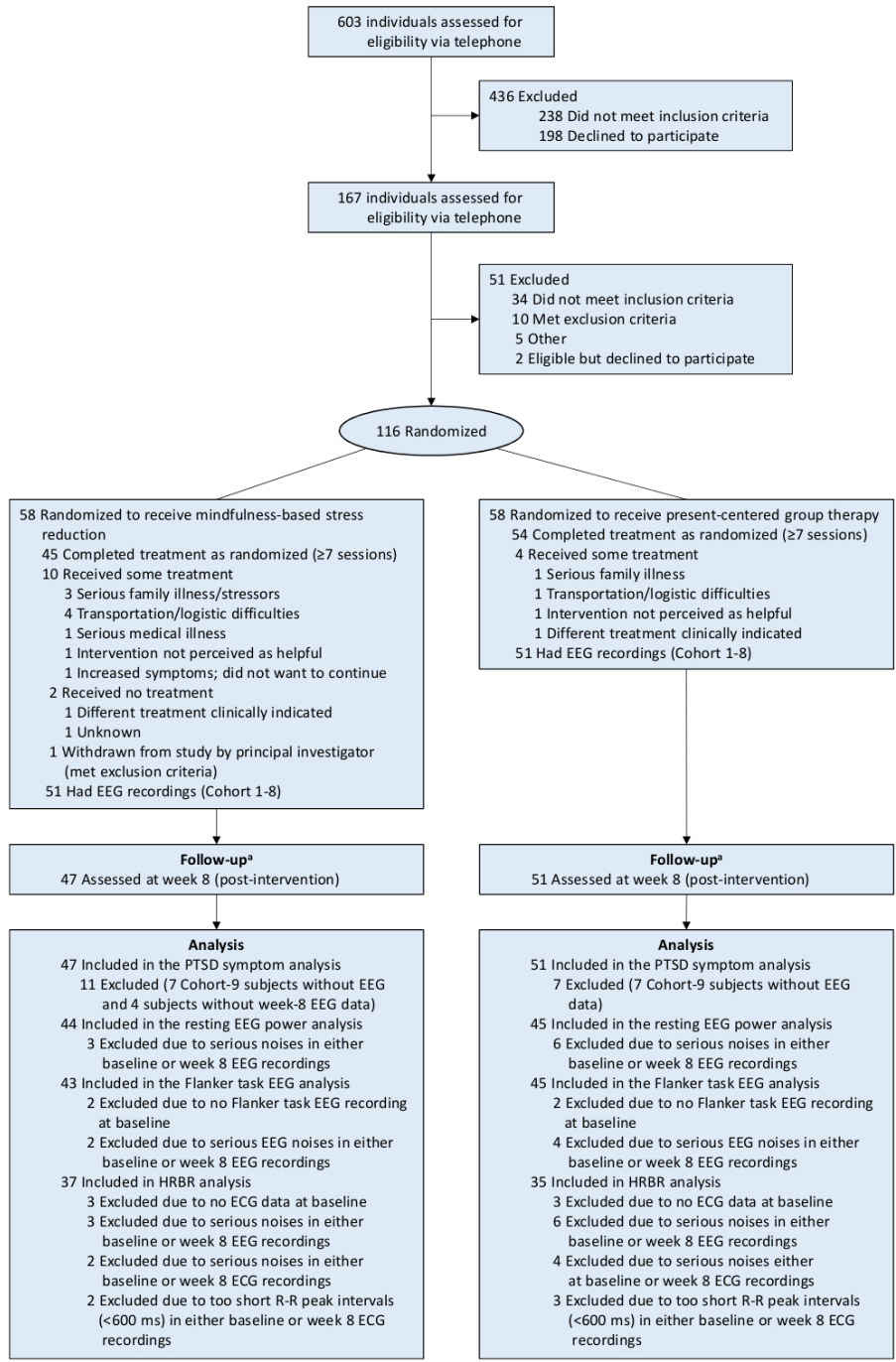

**Figure 2. Flow of Participants Through a Trial of MBSR vs. PCGT for Treatment of PTSD.** a. Reasons for loss to follow-up are unknown.

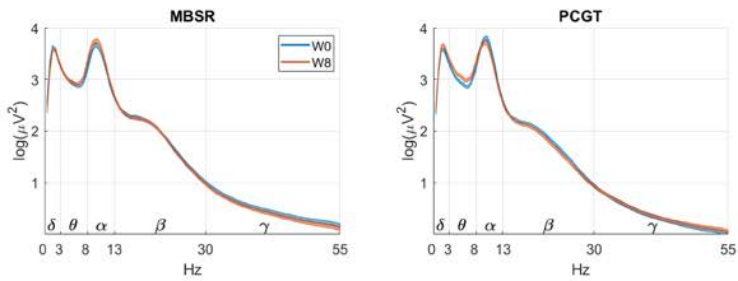
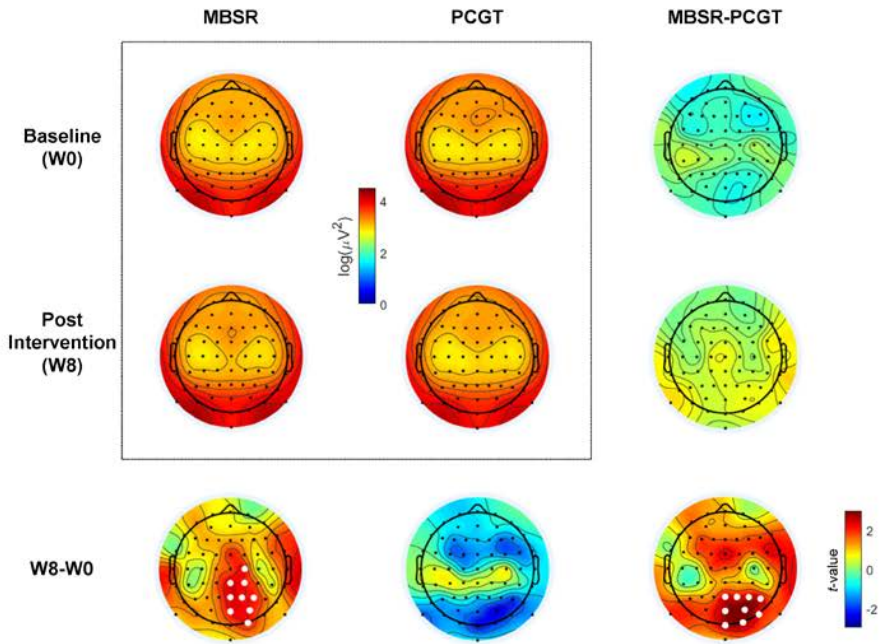
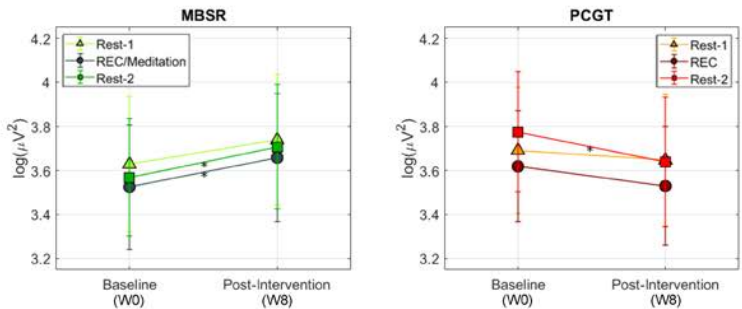

**Figure 3. Intervention-Related Changes in Spontaneous EEG Oscillatory Power.**
**A.** Grand average log spectral power (Mean ± 95% CI) of spontaneous EEG recordings collected from MBSR and PCGT participants at the baseline (week-0 [W0]) and post-intervention (week-8 [W8]). After intervention, the two groups had changes in the theta (3-7 Hz) and alpha (8-13 Hz) frequency power, which are not clearly discernible in the grand average spectral power due to the differences varying across electrodes. A trend-level theta power increase was observed after intervention (time effect: $F[1,87]=3.25$, $p=.075$, $\eta^2=.036$), which was not significantly different between groups (group-by-time interaction effect: $F[1,87]=1.46$, $p=.230$; see eFigure 1). There was a trend-level group-by-time interaction effect in alpha power ($F[1,87]=3.78$, $p=.055$, $\eta^2=.042$), which was characterized by the opposite directional trends of the intervention-related changes (increase in MBSR vs. decrease in PCGT). **B.** Topographical distributions of the spontaneous log alpha power and the *t*-statistics of the W8-W0 and MBSR-PCGT contrasts. Electrode cluster-based permutation tests identified post-intervention alpha power increases in the centroparietal electrode cluster (marked in white circles) in MBSR group ($p=.026$, $d=.33$). PCGT group had some alpha power decrease in the occipital electrodes, but no significant electrode cluster (3 or more adjacent electrodes with significant differences) was found. Group differences in the W8-W0 contrast of the alpha power (i.e., group-by-time interaction effects) were significant in the right posterior electrode cluster ($p=.005$, $d=.55$). **C.** The mean posterior alpha power of the electrodes showing the group-by-time interaction effects (the white electrodes of the bottom right topography) in the 3 states (Mean ± 95% CI). A follow-up RM-ANOVA with the W8-W0 difference mean posterior alpha power found that state ($F[2,174]=.26$, $p=.773$) and group-by-state effects ($F[2,174]=.97$, $p=.382$) were not significant, suggesting that each group had the similar intervention-related changes across the 3 states (* indicates a significant difference with $p<.05$).

## HEBR
Heartbeat-related EEG power TFDs showed strong wide-frequency and 2-7 Hz power increase around the ECG R- and T-peaks respectively, indicating large influences of cardiac field artifacts (Figure 5B). In the pre-

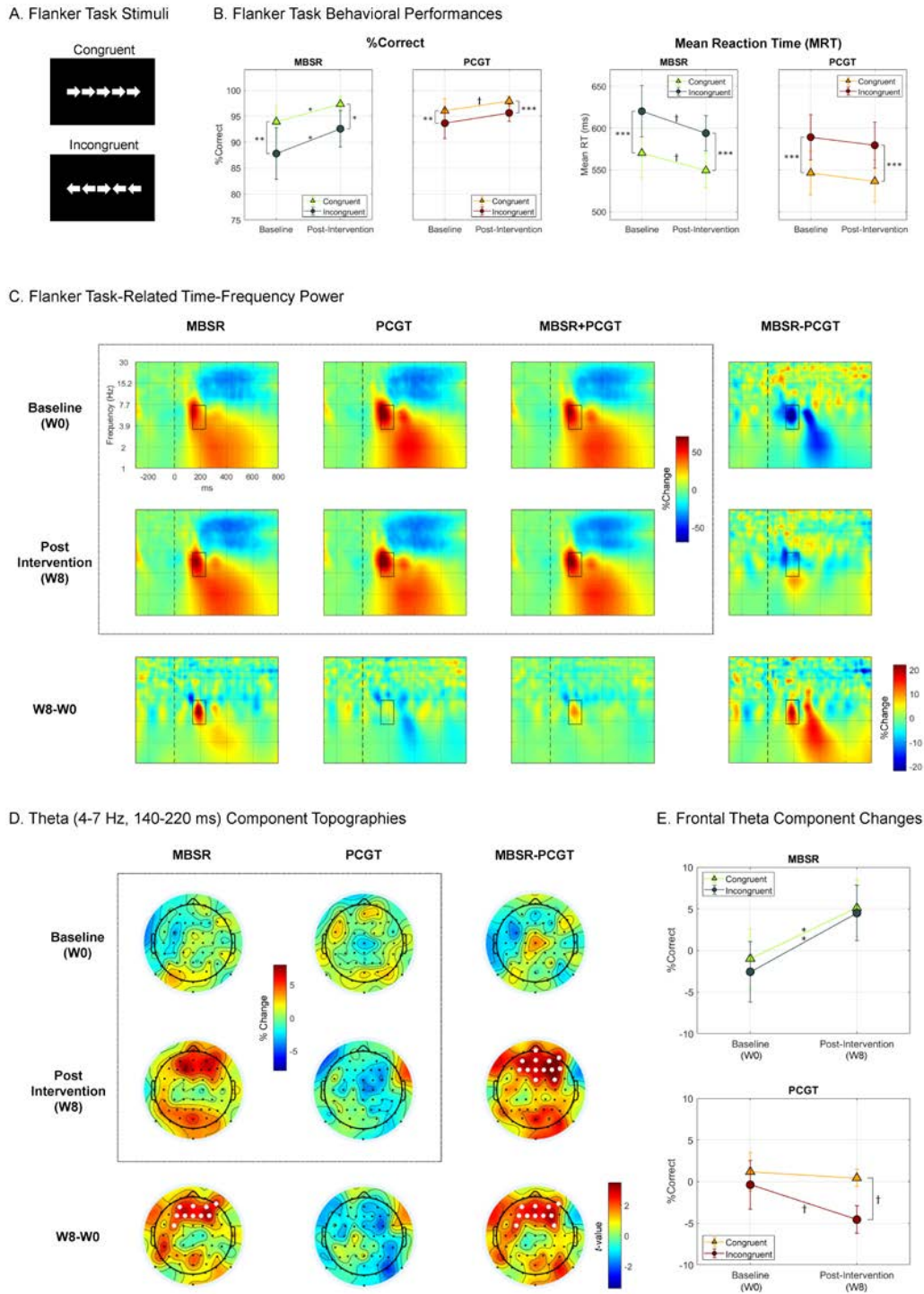

**Figure 4. Intervention-Related Changes in the Flanker Cognitive Task Performances and the Neural Oscillatory Responses.** **A.** Visual stimuli of the two conditions (congruent and incongruent) that were pseudo-randomly presented across trials (ITI: 1.8-2.2 sec) in the flanker task. Participants were required to respond by pressing a left or right button indicated by the central arrow regardless of the directions of the flanker arrows. **B.** Behavioral performance indices (Mean ± 95% CI) of %Correct and mean reaction time (MRT). Task performances were improved after intervention regardless of the task condition, especially in MBSR group. PCGT group had trend-level improvements in %Correct (congruent: $t[44]=2.01$, $p=.051$, $d=.30$; incongruent: $t[44]=-1.80$, $p=.078$, $d=.27$) but no significant changes in MRT, while MBSR group had a significant %Correct improvements (congruent: $t[42]=2.74$, $p=.009$, $d=.42$; incongruent: $t[42]=3.02$, $p=.004$, $d=.46$) and trend-level reductions in MRT (congruent: $t[42]=-1.68$, $p<.100$, $d=-.26$; incongruent: $t[42]=1.96$, $p=.057$, $d=-.30$). **C.** The grand average time-frequency (TF) power of EEG oscillatory activities time-locked to the onset of the task stimuli (collapsed across the conditions). The TF-power was normalized using the -300-0 ms pre-stimulus baseline TF-power and quantified as the %change relative to the baseline. Flanker task-related power TFDs were characterized by early theta power increases, subsequent alpha-beta power decreases, and delta power increases compared to the pre-stimulus baseline. The pre- vs. post-intervention difference TFD in the combined group (bottom row, third column) shows a post-intervention increase in 4-7 Hz theta power in 140-220 ms post-stimulus (highlighted by black rectangles). Only MBSR group had increases in the early theta power after intervention. **D.** Topographical distribution of the early theta-power and the *t*-statistics of the W8-W0 and MBSR-PCGT contrasts. Electrode cluster-based permutation tests found that MBSR group had the early theta-power increase in the frontocentral electrode cluster (marked in white circles; $p<.001$, $d=.39$), while PCGT group had no significant change in any electrode cluster. The theta power group difference was not significant in the baseline, but MBSR group had larger theta power than PCGT group in the frontal electrode cluster at post-intervention ($p<10^{-4}$, $d=.67$). Group difference in the W8-W0 contrast of the theta power was observed in the similar frontal electrode cluster ($p<.001$, $d=.54$). **E.** The mean frontal theta power of the electrodes showing the group-by-time interaction effects (the white electrodes of the bottom right topography) in the 2 conditions (Mean ± 95% CI). A follow-up RM-ANOVA with the W8-W0 difference mean frontal theta power found that condition ($F[2,86]=.17$, $p=.603$) and group-by-condition interaction ($F[2,174]=.87$, $p=.355$) effects were not significant, suggesting that each group had the similar intervention-related changes between the 2 conditions (* and † indicate a significant and a trend-level difference with $p<.05$ and $p<.10$, respectively).

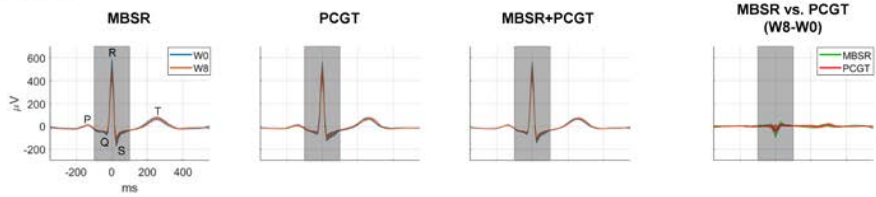
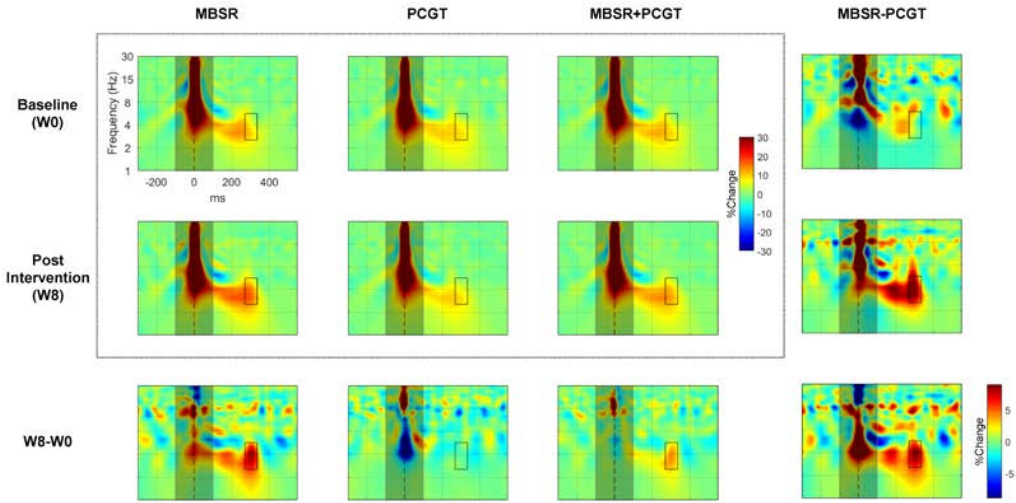
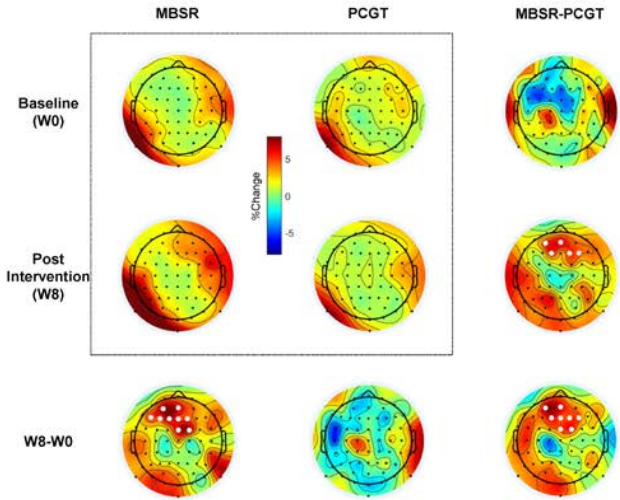
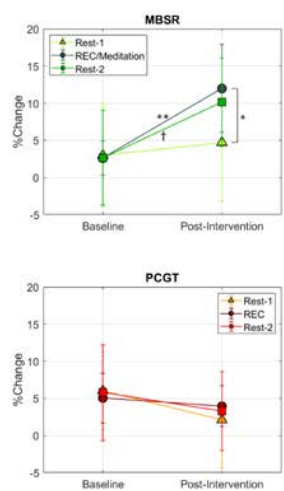

**Figure 5. Intervention-Related Changes in the Heartbeat-Evoked Brain Responses (HEBR).**
**A.** Grand average ECG waveforms time-locked to the R-peaks (collapsed across the 3 states) at pre- (W0) and post-intervention (W8) in MBSR, PCGT, and the combined group, in which P, Q, R, S, and T peak components of ECG are clearly shown. The W8-W0 contrast (right) shows substantial amplitude differences between the pre- and post-intervention ECG in the Q-R-S complex period (-100 to 100 ms), which were highlighted with gray shades (see also ECG power time-frequency distributions in eFigure 2). **B.** The grand average time-frequency (TF) power of EEG oscillatory activities time-locked to the ECG R-peaks. To quantify the heartbeat-related transient changes in the oscillatory activities, the TF-power was normalized using the -300 to -100 ms pre-R peak period, avoiding influences of strong artifacts around the Q-R-S complex (-100~100 ms) that were substantially different between the pre- and post-intervention ECGs. Large wide-frequency power increase around the R-peak and 2-7 Hz power increase around the T-peak in the heartbeat-related EEG power TFDs indicated strong influences of cardiac field artifacts in the HEBR measurements. In the combined group pre- vs. post-intervention difference TFD (bottom row, third column) that was minimally affected by the cardiac field artifacts, a post-intervention increase in 3-5 Hz theta frequency power was observed in 265-328 ms post-R peak (highlighted by black rectangles). Only MBSR group had increases in the theta HEBR after intervention. **C.** Topographical distribution of the theta HEBR and the *t*-statistics of the W8-W0 and MBSR-PCGT contrasts. HEBR topographies (circumscribed by a dotted rectangle) showed traces of the ECG T-peak component that was temporally adjacent to the theta HEBR and strongest in the marginal areas of the right fronto-temporal and the left occipito-temporal sites. The W8-W0 contrasts found that only MBSR group had significant theta HEBR increases in the frontal electrode cluster (marked in white circles; $p<.001$, $d=.46$), while PCGT had no significant changes. Group difference analyses found that the theta HEBR tended to be smaller in MBSR than in PCGT group at baseline, but no electrode cluster was found to be significant. After intervention, however, the theta HEBR was larger in MBSR than in PCGT group in the frontal electrode cluster ($p=.002$, $d=.57$). Group difference in the W8-W0 contrast of the theta HEBR was observed in the similar frontal electrode cluster ($p<.001$, $d=.61$). **D.** The mean frontal theta HEBR of the electrodes showing the group-by-time interaction effects (the white electrodes of the bottom right topography) in the 3 states (Mean ± 95% CI). A follow-up RM-ANOVA with the W8-W0 difference mean frontal theta HEBR power found a trend-level state effect ($F[2,140]=2.84$, $p=.062$, $\eta^2=.039$) and an insignificant group-by-state interaction effect ($F[2,140]=.60$, $p=.551$), suggesting that the intervention-related frontal theta HEBR changes were different between the states. Follow-up *t*-tests found that MBSR group had a significant and a trend-level increase of the frontal theta HEBR in Meditation ($t[36]=3.42$, $p=.002$, $d=.56$) and Rest-2 ($t[36]=1.79$, $p=.081$, $d=.30$) respectively, but not in Rest-1 ($t[36]=.36$, $p=.721$, $d=.06$). Also, the post-intervention theta HEBR was significantly larger in Meditation than in Rest-1 ($t[36]=2.44$, $p=.020$, $d=.41$). In contrast, PCGT group had no significant differences in any state and time contrasts. These results suggest that the MBSR-related frontal theta HEBR increase was state-dependent.

vs. post-intervention difference TFDs that were minimally affected by ECG artifacts, substantial TF-power increase was observed in 3-5 Hz and 265-328 ms after R-peak in the combined group. RM-ANOVA of the theta HEBR found significant group-by-time interaction ($F[1,70]=4.16$, $p=.045$, $\eta^2=.056$), group-by-electrode interaction ($F[63, 363.1]=2.36$, $p=.038$, $\eta^2=.033$), and state-by-electrode interaction ($F[126, 2560.1]=2.04$, $p<.001$, $\eta^2=.028$) effects. Only MBSR group had a significant increase in the theta HEBR in the frontal electrode cluster ($p<.001$, $d=.46$). The group differences in the intervention-related theta HEBR changes were significant in the similar frontal electrode cluster ($p<.001$, $d=.61$; Figure 5C). A follow-up RM-ANOVA with the post-pre intervention differences in the mean theta HEBR of the frontal electrodes found a trend-level state effect ($F[2,140]=2.84$, $p=.062$, $\eta^2=.039$). Follow-up $t$-tests found that unlike PCGT group having no differences in any time and state contrasts, MBSR group had a significant and a trend-level increase of the theta HEBR in Meditation ($t[36]=3.42$, $p=.002$, $d=.56$) and Rest-2 ($t[36]=1.79$, $p=.081$, $d=.30$) respectively, but not in Rest-1 ($t[36]=.36$, $p=.721$, $d=.06$). Also, the post-intervention theta HEBR was significantly larger in Meditation than in Rest-1 ($t[36]=2.44$, $p=.020$, $d=.41$; Figure 5D).

**Causal Mediation Effects of the EEG Measures**
The LDS models included mediators of the EEG measures that were quantified as the mean of the electrode clusters showing the significant group-by-time interaction effects. We found that the posterior alpha power (indirect path coefficient [$a*b$]=.016; 95% CI: -.024 to .075) and the task-related frontal theta response ($a*b$=-.029; 95% CI: -.091 to .024) had no significant mediation effect. However, the frontal theta HEBR during meditation had a significant mediation effect ($a*b$=-.040; 95% CI: -.092 to -.005; Figure 6A). Also, the direct effect of treatment type on the latent variable of PTSD symptom change was significant in the LDS model without the HEBR mediator ($c$=-.171; 95% CI: -.317 to -.033) but became insignificant after including the mediator in the model ($c'$=-.12; 95% CI: -.278 to.022). These results indicated that MBSR reduced PTSD symptom severity through improving the frontal theta HEBR.

**Brain Sources of the Theta HEBR Changes Predicting PTSD Symptom Improvements**
Nonparametric Spearman correlation analysis identified three cortical regions, whose estimated theta HEBR increases predicted the PTSD symptom improvements measured by PCL-sum score (Figure 6B), including the left ventral anterior cingulate cortex (L-vACC [MNI coordinates: -6,32, -12]: $\rho$=-.50, $p<.10^{-5}$), the right anterior insular cortex and claustrum (R-AIC/Cl [33,11,4]: $\rho$=-.39, $p<.001$), and the right lateral prefrontal cortex (R-LPFC [41,42,21]: $\rho$=-.50, $p<.10^{-5}$).

## DISCUSSION
Consistent with the literature and our hypothesis, we found that 8-week MBSR intervention improved the three neural activity measures of interest while the control intervention did not. First, MBSR increased posterior alpha power across the three states spontaneous EEG. Second, MBSR improved Flanker task performances and increased the task-related frontal theta responses regardless of the task conditions. Third, the frontal theta HEBR was increased in MBSR group especially during meditation and the resting after meditation (Rest-2). The LDS modeling based mediation analyses found that only the frontal theta HEBR significantly mediated the therapeutic effect of MBSR. The brain source level analysis identified that the theta HEBR increases in the L-vACC, R-AIC/Cl, and R-mPFC predicted the PTSD symptom reductions.

**Spontaneous EEG Alpha Power**
The intervention-related alpha power increase was observed only in MBSR group, but theta power increase was observed in both groups. Therefore, alpha but not theta frequency spontaneous brain activity increase was MBSR-specific. Drowsiness that increases delta~theta activities could affect the results, but the fatigue/drowsiness ratings were not significantly different between the pre- and post-intervention EEG sessions and between groups (eTable 2). Although there are debates on the functional significance of alpha oscillations, meditation-related alpha activity increase might reflect relaxed mental states with internally oriented attention[59] or enhanced somatosensory attention (body scanning and observing breathing sensations) that filters external sensory input[60]. Multiple EEG studies observed larger alpha power in meditation than in resting states[61–64]. However, the findings came from cross-sectional studies with long-term practitioners or novice participants instructed to practice simple meditative activity (e.g., breath counting), not reflecting meditation-related improvements from baseline. The present longitudinal study found that MBSR group had similar alpha power increases across the three states, suggest that the effect reflects trait changes rather than meditation state-specific changes.

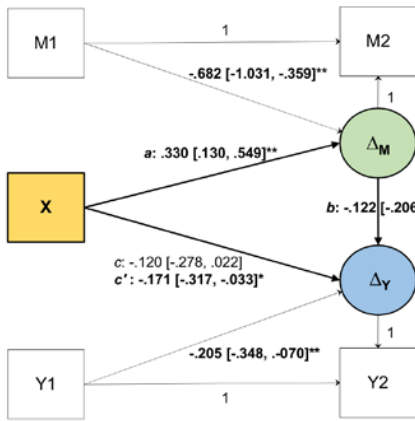

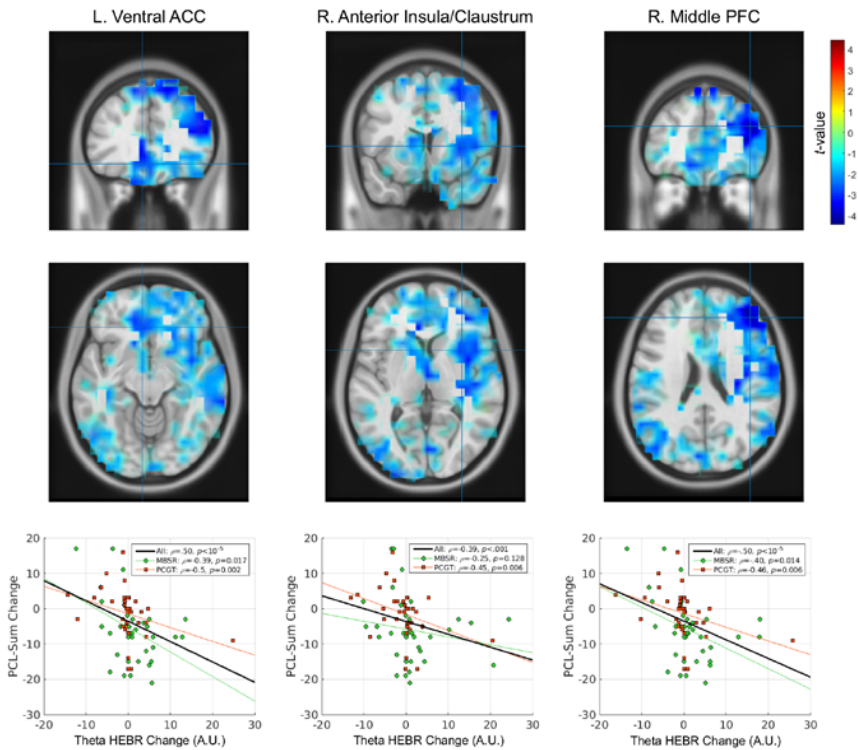

**Figure 6. Mediation Analyses and Brain Sources Whose HEBR Predicted PTSD Symptom Reductions. A.** Mediation analyses using latent difference score (LDS) modeling that tested if the heartbeat-related theta response (HEBR) mediated the therapeutic effect of MBSR (left: the LDS model path diagram with the path coefficients and their 95% confidence intervals in the brackets; right: the descriptions of the variables and the direct and indirect effects in the models). The indirect path coefficient was significant ($a*b$=-.040; 95% CI: -.092 to -.005), which represents the indirect effect of treatment-type ($X$) influencing the changes in the latent variable of PTSD symptom ($\Delta_Y$) through the changes in the latent variable of the frontal theta HEBR ($\Delta_M$). Also, the direct path coefficient from $X$ to $\Delta_Y$ was significant only in the LDS model without the theta HEBR mediator, suggesting the causal mediation effect of the theta HEBR change through which MBSR intervention exerted its therapeutic effect. The results of the LDS modeling with the other EEG mediators (the spontaneous alpha power and the task-related theta response) are summarized in eTable 3. The M-plus scripts of the LDS modeling are also available in eAppendix. **B.** Brain sources whose theta HEBR changes predicted PTSD symptom reductions. We conducted beamforming-based dynamic imaging of coherent sources (DICS)[55] analysis using Fieldtrip Matlab toolbox[56] to estimate oscillatory brain source activities of the theta HEBR (see the details in eAppendix). We conducted non-parametric Spearman correlation analyses between the post-intervention minus baseline difference (W8-W0) of PCL sum score and the W8-W0 difference of the theta HEBR power estimates in the standard MNI space. Significant negative correlations were observed, especially in the 3 brain regions; the left ventral anterior cingulate cortex (L-vACC [MNI coordinate: -6,32,-12]: $\rho$=-.50, $p$<.$10^{-5}$), the right anterior insular cortex and the claustrum (R-AIC/Cl [33,11,4]: $\rho$=-.39, $p$<.001), and the right lateral prefrontal cortex (R-LPFC [41,42,21]: $\rho$=-.50, $p$<.$10^{-5}$). As shown in the scatter plots, both groups had similar negative correlations. These results indicated that the theta HEBR increases in the brain regions related to interoception and attentional control predicted PTSD symptom reductions.

## Cognitive Task-Related Theta Responses

MBSR tended to improve the task performances more than PCGT given that MBSR group had significant and trend level improvements in %Correct and MRT respectively for the both task conditions, while PCGT group had only a trend level %Correct improvements for congruent trials. Consistently, MBSR significantly increased the task-related frontal theta oscillatory power regardless of the task conditions, while PCGT group had slight reductions in the theta component for incongruent trials. EEG and MEG studies have identified flanker task-related frontal theta responses that reflect top-down attentional control processes of ACC and lateral PFC for resolving response conflicts[33,65]. Although previous cross-sectional studies found long-term mindfulness meditators had higher performance accuracy[66–68] and larger frontal theta neural synchrony in flanker-type tasks than non-meditators[68], to our knowledge, the current findings represent the first longitudinal demonstrations that mindfulness meditation enhances top-down neural processes for attention-control.

**Theta HEBR**

HEBR has been mostly studied in time-domain (i.e., heartbeat-evoked potential [HEP]), and studies using electrocorticogram (ECoG) that is minimally affected by cardiac field artifacts found phasic HEP response around 300 ms after ECG R-peaks in the somatosensory cortex[41] and insula[69] during resting state. Also, Park et al identified the strongest R-peak related TF neural oscillatory synchrony of ECoG in 4-7 Hz around 200 ms post R-peaks especially in the insula[69]. As studies have shown that bodily self-consciousness is associated with the HEP amplitude around 300 ms after the ECG R-peak over fronto-central regions[27,69], our findings of the MBSR-related increase in the frontal theta HEBR (3-5 Hz, 265-328 ms post-R-peak) might represent enhanced cerebral interoceptive functions promoting bodily awareness. MBSR includes practices attending to the present-moment bodily sensations (body scanning, mindful observation of breathing, and yoga), which appear to improve the brain's general interoceptive capacity responding to subtle internal bodily sensations like heartbeats. MBSR group had the significant frontal theta HEBR increase during Meditation but not in Rest-1, suggesting that the interoceptive brain response increase was Meditation state-dependent. Given the trend-level increase of the HEBR observed in Rest-2, the interoceptive meditation effect appears to be sustained for a while in a somewhat diminished degree. Importantly, cardiac parameters including heart rate and heart rate variability (eTable 2), ECG amplitude (Figure 5A), and the original and normalized TF-power of ECG signals (eFigure 2) in the theta HEBR TF-window were not different between pre- and post-intervention and between groups. Therefore, the HEBR changes were not driven by ECG parameters and artifacts.

It is noteworthy that the insula activation was the most consistent finding in not only a meta-analysis of fMRI studies of cardiac interoceptive attention[70], but also systematic review of longitudinal fMRI studies of meditation[71]. Also, a recent fMRI study using a task employing interoceptive attention toward breathing rhythm found that the right AIC played the central role in respiratory interoceptive attention[72]. A supplementary region-of-interest (ROI) analysis using the MNI coordinate of AIC [28,28,0][72] found a significant group-by-time interaction effect of the AIC theta HEBR ($F_{[1,70]}=2.31$, $p=.021$, $\eta^2=.074$), indicating that only MBSR group had significant increase in the AIC theta HEBR ($t_{[36]}=2.40$, $p=.022$; $d=.39$, eFigure 3). These results suggest that MBSR improved the cerebral interoceptive functions especially in the AIC.

**Causal Mediation Effects of the Frontal Theta HEBR**

The frontal theta HEBR increase had a significant causal mediation effect in the LDS model, demonstrating that MBSR exerted its therapeutic effects through improvements in interoceptive brain mechanisms. Supplementary mediation analyses for the 3 sub-symptoms of PTSD found that the HEBR had mediation effects, especially for re-experiencing symptoms (eTable 4). The brain source-level correlational analysis indicated that the improvements in PTSD symptoms were predicted by the theta HEBR increases in ACC and AIC that are known as the brain sources of HEBR[39] and LPFC that plays the central role in attention-control[73,74]. Given the fMRI finding that in-the-moment re-experiencing symptoms of PTSD were associated with abnormal activations in insula, ACC, and inferior PFC[75], mindfulness meditation promoting interoceptive attention to present-moment might normalize the dysregulated interoceptive and prefrontal functions and improve intrusion symptoms associated with trauma. Given the significant MBSR-related increases but non-significant mediation effects of the spontaneous alpha power and the attentional task-related theta responses, they might reflect beneficial effects of MBSR that did not critically contribute to the therapeutic effect of MBSR.

**Translational and Clinical Implications and Caveats**

PTSD symptoms are associated with over-reactivity of the sympathetic nervous system[76,77] and failures of the central nervous system (CNS) in regulating the autonomic nervous system (ANS)[78–80]. The efficacy of mindfulness meditation in regulating ANS and improving ANS-CNS interactions[81] might primarily originate from practicing the present-moment attention to internal bodily sensations that enhance interoceptive brain function, which appears to be the essential therapeutic mechanism of MBSR. This implies the possibility of optimizing the current practices of meditative interventions for various stress-related mental disorders by maximizing the efficacy to improve the cerebral interoceptive mechanism.

Some limitations of this study should be noted. First, although participants in the two intervention groups were not significantly different in any demographic variables and clinical and EEG measures at baseline, we cannot rule out that some confounding factors not included in the mediation analyses might partially influence the results. Second, due to the limited number of daily treatment adherence measurements, we were unable to assess the relationship between the homework MBSR practice hours and the changes in the clinical and EEG measures. Third, given the limited spatial resolution of the present source localization method, we cannot conclude that the theta HEBR is increased in exactly the same region of the cortex of the response.

## CONCLUSIONS

Although mindfulness meditation had multiple beneficial effects on the brain functions, the cerebral interoceptive function was found to be the core therapeutic mechanism of MBSR, which appears to improve the homeostatic CNS-ANS interactions and emotional regulation that are disturbed in people with PTSD.

## ACKNOWLEDGEMENT

**Author Contributions**: Dr Kang had full access to all of the data in the study and takes responsibility for the integrity of the data and the accuracy of the data analysis.
*Study concept and design*: Kang, Sponheim, Lim.
*Acquisition, analysis, or interpretation of data*: Kang, Sponheim, Lim.
*Drafting of the manuscript*: Kang.
*Critical revision of the manuscript for important intellectual content*: All authors.
*Statistical analysis*: Kang.
*Obtained funding*: Lim.
*Administrative, technical, or material support*: Kang, Spoheim, Lim.
*Study supervision*: Kang, Sponheim, Lim.
**Conflict of Interest Disclosures**: All authors have no conflicts of interest to report.
**Funding/Support**: This material is the result of work supported with resources and the use of facilities at the Minneapolis VA Health Care System, Minneapolis, Minnesota. This research was supported by VA grant 5I01CX000683-01 to Dr Lim.
**Role of the Funder/Sponsor**: The funder had no role in the design and conduct of the study; collection, management, analysis, and interpretation of the data; preparation, review, or approval of the manuscript; or decision to submit the manuscript for publication.
**Disclaimer**: The views expressed in this article are those of the authors and do not reflect the official policy or position of the VA.
**Additional Contributions**:
Minneapolis VA Health Care System staff clinical psychologists Melissa Polusny, PhD, Christopher Erbes, PhD, Greg Lamberty, PhD, and John Rodman, PhD, contributed to the design and supervision of the clinical trial as part of their of research duties.
Minneapolis VA Health Care System statistician Paul Thuras, PhD, staff clinical psychologist Rose C. Collins, and research staff Amy Moran, MA contributed to clinical data acquisition as part of their research and clinical duties.
Minneapolis VA Health Care System EEG technician Abraham C. Van Voorhis contributed to EEG data collection and was provided compensation for his role in the study.
Minneapolis VA Health Care System clinicians Torricia Yamada, PhD, Carolyn Anderson, PhD, Maureen Kennedy, PsyD, Kelly Petska, PhD, Jacqueline Wright, LICSW, Nancy Koets, PsyD, Margaret Gavian, PhD, and Ivy Miller, PhD, contributed to intervention delivery as part of their provision of clinical care.
Mariann Johnson, BA, University of Minnesota Center for Spirituality and Healing, contributed to intervention delivery and was provided compensation for her role in the study.
Terry Pearson, RPh, MBA, University of Minnesota Center for Spirituality and Healing, provided consultation on mindfulness-based stress reduction and evaluation of treatment fidelity and was provided compensation for her role in the study.
Melissa Wattenberg, PhD, VA Boston Healthcare System and Boston University School of Medicine, provided training and consultation on present-centered group therapy, for which she received no compensation.

Leah Gause, MA, and Cassandra Sartor, MA, Minneapolis VA Health Care System, served as independent assessors and were provided compensation for their roles in the study.

Doris Clancy, MA, and Cory Voecks, MA, provided administrative support and were provided compensation for their roles in the study.

Elizabeth Gibson, BA, Minneapolis VA Health Care System, provided editing assistance and received no compensation.

**REFERENCES**

1. Tang Y-Y, Hölzel BK, Posner MI. The neuroscience of mindfulness meditation. *Nat Rev Neurosci*. 2015;16(4):213-225. doi:10.1038/nrn3916
2. Dahl CJ, Lutz A, Davidson RJ. Reconstructing and deconstructing the self: cognitive mechanisms in meditation practice. *Trends Cogn Sci*. 2015;19(9):515-523. doi:10.1016/J.TICS.2015.07.001
3. Sedlmeier P, Eberth J, Schwarz M, et al. The psychological effects of meditation: A meta-analysis. *Psychol Bull*. 2012;138(6):1139-1171. doi:10.1037/a0028168
4. Kabat-Zinn J. *Full Catastrophe Living : Using the Wisdom of Your Body and Mind to Face Stress, Pain, and Illness*. New York, NY: Random House Publishing Group; 1990.
5. Baer RA. Mindfulness Training as a Clinical Intervention: A Conceptual and Empirical Review. *Clin Psychol Sci Pract*. 2006;10(2):125-143. doi:10.1093/clipsy.bpg015
6. Kim SH, Schneider SM, Kravitz L, Mermier C, Burge MR. Mind-body practices for posttraumatic stress disorder. *J Investig Med*. 2013;61(5):827-834. doi:10.2310/JIM.0b013e3182906862
7. Polusny MA, Erbes CR, Thuras P, et al. Mindfulness-Based Stress Reduction for Posttraumatic Stress Disorder Among Veterans. *JAMA*. 2015;314(5):456-465. doi:10.1001/jama.2015.8361
8. Heffner KL, Crean HF, Kemp JE. Meditation programs for veterans with posttraumatic stress disorder: Aggregate findings from a multi-site evaluation. *Psychol Trauma*. 2016;8(3):365-374. doi:10.1037/tra0000106
9. Fox KCR, Nijeboer S, Dixon ML, et al. Is meditation associated with altered brain structure? A systematic review and meta-analysis of morphometric neuroimaging in meditation practitioners. *Neurosci Biobehav Rev*. 2014;43:48-73. doi:10.1016/J.NEUBIOREV.2014.03.016
10. Cahn BR, Polich J. Meditation states and traits: EEG, ERP, and neuroimaging studies. *Psychol Bull*. 2006;132(2):180-211. doi:10.1037/0033-2909.132.2.180
11. Lomas T, Ivtzan I, Fu CHY. A systematic review of the neurophysiology of mindfulness on EEG oscillations. *Neurosci Biobehav Rev*. 2015;57:401-410. doi:10.1016/j.neubiorev.2015.09.018
12. Critchley HD, Garfinkel SN. Interoception and emotion. *Curr Opin Psychol*. 2017;17:7-14. doi:10.1016/J.COPSYC.2017.04.020
13. Khalsa SS, Adolphs R, Cameron OG, et al. Interoception and Mental Health: A Roadmap. *Biol psychiatry Cogn Neurosci neuroimaging*. 2018;3(6):501-513. doi:10.1016/j.bpsc.2017.12.004
14. Farb NAS, Segal Z V., Anderson AK. Mindfulness meditation training alters cortical representations of interoceptive attention. *Soc Cogn Affect Neurosci*. 2013;8(1):15-26. doi:10.1093/scan/nss066
15. Farb N, Daubenmier J, Price CJ, et al. Interoception, contemplative practice, and health. *Front Psychol*. 2015;6:763. doi:10.3389/fpsyg.2015.00763
16. Gibson J. Mindfulness, Interoception, and the Body: A Contemporary Perspective. *Front Psychol*. 2019;10:2012. doi:10.3389/fpsyg.2019.02012
17. Solomonova E, Fox KCR, Nielsen T. Methodological considerations for the neurophenomenology of dreaming: commentary on Windt's "Reporting dream experience". *Front Hum Neurosci*. 2014;8:317. doi:10.3389/fnhum.2014.00317
18. Strigo IA, Craig ADB. Interoception, homeostatic emotions and sympathovagal balance. *Philos Trans R Soc Lond B Biol Sci*. 2016;371(1708). doi:10.1098/rstb.2016.0010
19. Daubenmier J, Sze J, Kerr CE, Kemeny ME, Mehling W. Follow your breath: respiratory interoceptive accuracy in experienced meditators. *Psychophysiology*. 2013;50(8):777-789. doi:10.1111/psyp.12057
20. Schandry R, Montoya P. Event-related brain potentials and the processing of cardiac activity. *Biol Psychol*. 1996;42(1-2):75-85. doi:10.1016/0301-0511(95)05147-3
21. Montoya P, Schandry R, Müller A. Heartbeat evoked potentials (HEP): topography and influence of

cardiac awareness and focus of attention. *Electroencephalogr Clin Neurophysiol*. 1993;88(3):163-172. doi:10.1016/0168-5597(93)90001-6
22. Petzschner FH, Weber LA, Wellstein K V., Paolini G, Do CT, Stephan KE. Focus of attention modulates the heartbeat evoked potential. *Neuroimage*. 2019;186:595-606. doi:10.1016/J.NEUROIMAGE.2018.11.037
23. Kim J, Park H-D, Kim KW, et al. Sad faces increase the heartbeat-associated interoceptive information flow within the salience network: a MEG study. *Sci Rep*. 2019;9(1):430. doi:10.1038/s41598-018-36498-7
24. Luft CDB, Bhattacharya J. Aroused with heart: Modulation of heartbeat evoked potential by arousal induction and its oscillatory correlates. *Sci Rep*. 2015;5(1):15717. doi:10.1038/srep15717
25. Gentsch A, Sel A, Marshall AC, Schütz-Bosbach S. Affective interoceptive inference: Evidence from heart-beat evoked brain potentials. *Hum Brain Mapp*. 2019;40(1):20-33. doi:10.1002/hbm.24352
26. Babo-Rebelo M, Richter CG, Tallon-Baudry C. Neural Responses to Heartbeats in the Default Network Encode the Self in Spontaneous Thoughts. *J Neurosci*. 2016;36(30). http://www.jneurosci.org/content/36/30/7829. Accessed August 27, 2017.
27. Park H-D, Bernasconi F, Bello-Ruiz J, Pfeiffer C, Salomon R, Blanke O. Transient Modulations of Neural Responses to Heartbeats Covary with Bodily Self-Consciousness. *J Neurosci*. 2016;36(32):8453-8460. doi:10.1523/JNEUROSCI.0311-16.2016
28. Blevins CA, Weathers FW, Davis MT, Witte TK, Domino JL. The Posttraumatic Stress Disorder Checklist for *DSM-5* (PCL-5): Development and Initial Psychometric Evaluation. *J Trauma Stress*. 2015;28(6):489-498. doi:10.1002/jts.22059
29. Resick PA, Wachen JS, Mintz J, et al. A randomized clinical trial of group cognitive processing therapy compared with group present-centered therapy for PTSD among active duty military personnel. *J Consult Clin Psychol*. 2015;83(6):1058-1068. doi:10.1037/ccp0000016
30. Frost ND, Laska KM, Wampold BE. The evidence for present-centered therapy as a treatment for posttraumatic stress disorder. *J Trauma Stress*. 2014;27(1):1-8. doi:10.1002/jts.21881
31. Goyal M, Singh S, Sibinga EMS, et al. Meditation Programs for Psychological Stress and Well-being. *JAMA Intern Med*. 2014;174(3):357. doi:10.1001/jamainternmed.2013.13018
32. Metting van Rijn AC, Peper A, Grimbergen CA. High-quality recording of bioelectric events. *Med Biol Eng Comput*. 1990;28(5):389-397. doi:10.1007/BF02441961
33. McDermott TJ, Wiesman AI, Proskovec AL, Heinrichs-Graham E, Wilson TW. Spatiotemporal oscillatory dynamics of visual selective attention during a flanker task. *Neuroimage*. 2017;156:277-285. doi:10.1016/J.NEUROIMAGE.2017.05.014
34. Hyvärinen A. Fast and robust fixed-point algorithms for independent component analysis. *IEEE Trans Neural Networks*. 1999;10(3):626-634. doi:10.1109/72.761722
35. Hyvärinen A, Oja E. Independent component analysis: Algorithms and applications. *Neural Networks*. 2000;13(4-5):411-430. doi:10.1016/S0893-6080(00)00026-5
36. Delorme A, Sejnowski T, Makeig S. Enhanced detection of artifacts in EEG data using higher-order statistics and independent component analysis. *Neuroimage*. 2007;34(4):1443-1449. doi:10.1016/j.neuroimage.2006.11.004
37. Kang SS. Addressing Measurements Issues in Electroencephalography Studies of Meditations as Alternative Interventions of Post-Traumatic Stress Disorders. *Psychol Trauma*.
38. Rajan JJ, Rayner PJW. IEE proceedings. Vision, image, and signal processing. *IEE Proc - Vision, Image Signal Process*. 1997;144(2):116-123. http://digital-library.theiet.org/content/journals/10.1049/ip-vis_19971093. Accessed March 25, 2018.
39. Park H-D, Blanke O. Heartbeat-evoked cortical responses: Underlying mechanisms, functional roles, and methodological considerations. *Neuroimage*. April 2019. doi:10.1016/j.neuroimage.2019.04.081
40. Dirlich G, Vogl L, Plaschke M, Strian F. Cardiac field effects on the EEG. *Electroencephalogr Clin Neurophysiol*. 1997;102(4):307-315. doi:10.1016/S0013-4694(96)96506-2
41. Kern M, Aertsen A, Schulze-Bonhage A, Ball T. Heart cycle-related effects on event-related potentials, spectral power changes, and connectivity patterns in the human ECoG. *Neuroimage*. 2013;81:178-190. doi:10.1016/j.neuroimage.2013.05.042
42. Berger H. On the electroencephalogram of man. Second report. *Electroencephalogr Clin Neurophysiol*. 1969;Suppl 28:75+. http://www.ncbi.nlm.nih.gov/pubmed/4188919. Accessed July 29, 2020.

43. Cohen L, Leon. *Time-Frequency Analysis*. Englewood Cliffs, NJ.: Prentice Hall PTR; 1995.
44. Bernat EM, Williams WJ, Gehring WJ. Decomposing ERP time-frequency energy using PCA. *Clin Neurophysiol*. 2005;116(6):1314-1334. doi:10.1016/j.clinph.2005.01.019
45. Jeong J, Williams WJ. A new formulation of generalized discrete-time time-frequency distributions. In: *Proceedings of the IEEE ICASSP-91.* ; 1991:3189-3192.
46. Park H-D, Bernasconi F, Salomon R, et al. Neural Sources and Underlying Mechanisms of Neural Responses to Heartbeats, and their Role in Bodily Self-consciousness: An Intracranial EEG Study. *Cereb Cortex*. 2017;116:1-14. doi:10.1093/cercor/bhx136
47. Schandry R, Sparrer B, Weitkunat R. From the heart to the brain: a study of heartbeat contingent scalp potentials. *Int J Neurosci*. 1986;30(4):261-275. http://www.ncbi.nlm.nih.gov/pubmed/3793380. Accessed October 26, 2018.
48. Kriegeskorte N, Simmons WK, Bellgowan PSF, Baker CI. Circular analysis in systems neuroscience: the dangers of double dipping. *Nat Neurosci*. 2009;12(5):535-540. doi:10.1167/8.6.88
49. Huynh H, Feldt LS. Estimation of the Box Correction for Degrees of Freedom from Sample Data in Randomized Block and Split-Plot Designs. *J Educ Stat*. 1976;1(1):69-82. doi:10.3102/10769986001001069
50. Maris E, Oostenveld R. Nonparametric statistical testing of EEG- and MEG-data. *J Neurosci Methods*. 2007;164(1):177-190. doi:10.1016/j.jneumeth.2007.03.024
51. McArdle JJ, Hamagami F. Latent difference score structural models for linear dynamic analyses with incomplete longitudinal data. In: *New Methods for the Analysis of Change.* Decade of behavior. Washington, DC, US: American Psychological Association; 2001:139-175. doi:10.1037/10409-005
52. McArdle JJ, Nesselroade JR. Using multivariate data to structure developmental change. In: *Life-Span Developmental Psychology: Methodological Contributions.* The West Virginia University conferences on life-span developmental psychology. Hillsdale, NJ, US: Lawrence Erlbaum Associates, Inc; 1994:223-267.
53. Finch WH, Shim SS. A Comparison of Methods for Estimating Relationships in the Change Between Two Time Points for Latent Variables. *Educ Psychol Meas*. 2018;78(2):232-252. doi:10.1177/0013164416680701
54. Muthén LK, Muthén BO. *Mplus User's Guide*. Eighth Edi. Los Angeles, CA: Muthén & Muthén
55. Gross J, Kujala J, Hamalainen M, Timmermann L, Schnitzler A, Salmelin R. Dynamic imaging of coherent sources: Studying neural interactions in the human brain. *Proc Natl Acad Sci U S A*. 2001;98(2):694-699. doi:10.1073/pnas.98.2.694
56. Vorwerk J, Oostenveld R, Piastra MC, Magyari L, Wolters CH. The FieldTrip-SimBio pipeline for EEG forward solutions. *Biomed Eng Online*. 2018;17(1):37. doi:10.1186/s12938-018-0463-y
57. Manera AL, Dadar M, Fonov V, Collins DL. CerebrA, registration and manual label correction of Mindboggle-101 atlas for MNI-ICBM152 template. *Sci Data*. 2020;7(1):1-9. doi:10.1038/s41597-020-0557-9
58. Van Veen BD, Van Drongelen W, Yuchtman M, Suzuki A. Localization of brain electrical activity via linearly constrained minimum variance spatial filtering. *IEEE Trans Biomed Eng*. 1997;44(9):867-880. doi:10.1109/10.623056
59. Shaw JC. Intention as a component of the alpha-rhythm response to mental activity. *Int J Psychophysiol*. 1996;24(1-2):7-23. doi:10.1016/S0167-8760(96)00052-9
60. Kerr CE, Sacchet MD, Lazar SW, Moore CI, Jones SR. Mindfulness starts with the body: somatosensory attention and top-down modulation of cortical alpha rhythms in mindfulness meditation. *Front Hum Neurosci*. 2013;7:12. doi:10.3389/fnhum.2013.00012
61. Lagopoulos J, Xu J, Rasmussen I, et al. Increased Theta and Alpha EEG Activity During Nondirective Meditation. *J Altern Complement Med*. 2009;15(11):1187-1192. doi:10.1089/acm.2009.0113
62. Yu X, Fumoto M, Nakatani Y, et al. Activation of the anterior prefrontal cortex and serotonergic system is associated with improvements in mood and EEG changes induced by Zen meditation practice in novices. *Int J Psychophysiol*. 2011;80(2):103-111. doi:10.1016/J.IJPSYCHO.2011.02.004
63. Milz P, Faber PL, Lehmann D, Kochi K, Pascual-Marqui RD. sLORETA intracortical lagged coherence during breath counting in meditation-naïve participants. *Front Hum Neurosci*. 2014;8:303. doi:10.3389/fnhum.2014.00303
64. Kasamatsu A, Hirai T. AN ELECTROENCEPHALOGRAPHIC STUDY ON THE ZEN MEDITATION

(ZAZEN). *Psychiatry Clin Neurosci*. 1966;20(4):315-336. doi:10.1111/j.1440-1819.1966.tb02646.x
65. Nigbur R, Ivanova G, Stürmer B. Theta power as a marker for cognitive interference. *Clin Neurophysiol*. 2011;122(11):2185-2194. doi:10.1016/J.CLINPH.2011.03.030
66. van den Hurk PAM, Giommi F, Gielen SC, Speckens AEM, Barendregt HP. Greater efficiency in attentional processing related to mindfulness meditation. *Q J Exp Psychol (Hove)*. 2010;63(6):1168-1180. doi:10.1080/17470210903249365
67. Jo H-G, Schmidt S, Inacker E, Markowiak M, Hinterberger T. Meditation and attention: A controlled study on long-term meditators in behavioral performance and event-related potentials of attentional control. *Int J Psychophysiol*. 2016;99:33-39. doi:10.1016/J.IJPSYCHO.2015.11.016
68. Jo H-G, Malinowski P, Schmidt S. Frontal Theta Dynamics during Response Conflict in Long-Term Mindfulness Meditators. *Front Hum Neurosci*. 2017;11:299. doi:10.3389/fnhum.2017.00299
69. Park H-D, Bernasconi F, Salomon R, et al. Neural Sources and Underlying Mechanisms of Neural Responses to Heartbeats, and their Role in Bodily Self-consciousness: An Intracranial EEG Study. *Cereb Cortex*. 2018;28(7):2351-2364. doi:10.1093/cercor/bhx136
70. Schulz SM. Neural correlates of heart-focused interoception: a functional magnetic resonance imaging meta-analysis. *Philos Trans R Soc Lond B Biol Sci*. 2016;371(1708). doi:10.1098/rstb.2016.0018
71. Young KS, van der Velden AM, Craske MG, et al. The impact of mindfulness-based interventions on brain activity: A systematic review of functional magnetic resonance imaging studies. *Neurosci Biobehav Rev*. 2018;84:424-433. doi:10.1016/J.NEUBIOREV.2017.08.003
72. Wang X, Wu Q, Egan L, et al. Anterior insular cortex plays a critical role in interoceptive attention. *Elife*. 2019;8. doi:10.7554/eLife.42265
73. Braver TS. The variable nature of cognitive control: A dual mechanisms framework. *Trends Cogn Sci*. 2012;16(2):106-113. doi:10.1016/j.tics.2011.12.010
74. MacDonald AW, Cohen JD, Stenger VA, Carter CS. Dissociating the role of the dorsolateral prefrontal and anterior cingulate cortex in cognitive control. *Science*. 2000;288(5472):1835-1838. doi:10.1126/science.288.5472.1835
75. Hopper JW, Frewen PA, van der Kolk BA, Lanius RA. Neural correlates of reexperiencing, avoidance, and dissociation in PTSD: symptom dimensions and emotion dysregulation in responses to script-driven trauma imagery. *J Trauma Stress*. 2007;20(5):713-725. doi:10.1002/jts.20284
76. Brudey C, Park J, Wiaderkiewicz J, Kobayashi I, Mellman TA, Marvar PJ. Autonomic and inflammatory consequences of posttraumatic stress disorder and the link to cardiovascular disease. *Am J Physiol Regul Integr Comp Physiol*. 2015;309(4):R315-21. doi:10.1152/ajpregu.00343.2014
77. Blechert J, Michael T, Grossman P, Lajtman M, Wilhelm FH. Autonomic and Respiratory Characteristics of Posttraumatic Stress Disorder and Panic Disorder. *Psychosom Med*. 2007;69(9):935-943. doi:10.1097/PSY.0b013e31815a8f6b
78. Fenster RJ, Lebois LAM, Ressler KJ, Suh J. Brain circuit dysfunction in post-traumatic stress disorder: from mouse to man. *Nat Rev Neurosci*. 2018;19(9):535-551. doi:10.1038/s41583-018-0039-7
79. Sherin JE, Nemeroff CB. Post-traumatic stress disorder: the neurobiological impact of psychological trauma. *Dialogues Clin Neurosci*. 2011;13(3):263-278. http://www.ncbi.nlm.nih.gov/pubmed/22034143. Accessed July 27, 2020.
80. Williamson JB, Porges EC, Lamb DG, Porges SW. Maladaptive autonomic regulation in PTSD accelerates physiological aging. *Front Psychol*. 2014;5:1571. doi:10.3389/fpsyg.2014.01571
81. Tang Y-Y, Ma Y, Fan Y, et al. Central and autonomic nervous system interaction is altered by short-term meditation. *Proc Natl Acad Sci U S A*. 2009;106(22):8865-8870. doi:10.1073/pnas.0904031106